\documentclass[a4paper,11pt]{article}
\usepackage{nfraconf,graphicx,cite,psfig}
\def\gsim{\mathrel{\hbox{\rlap{\lower.55ex \hbox {$\sim$}}
                   \kern-.3em \raise.4ex \hbox{$>$}}}}

\begin{document}
\baselineskip=13pt

\title{The Unique Potential of SKA Radio Observations of Gamma-Ray Bursts.}

\author{T. J. Galama}
\address{Astronomy\\
Mail Stop 105-24, Robinson Lab\\
California Institute of Technology\\
Pasadena, CA 91125\\
U.S.A.\\
E-mail: tjg@astro.caltech.edu}

\author{A. G. De Bruyn}
\address{Netherlands Foundation for Research in Astronomy and
Kapteyn Astronomical Institute\\
E-mail: ger@nfra.nl}


\maketitle

\abstract{Radio observations with the Square Kilometer Array (SKA)
provide the agility, sensitivity, and spectral coverage to trace the
evolution of the size, shape and spectra of gamma-ray burst (GRB)
remnants from the earliest moments on. In the first hours to days
after the burst a major, and unique tool, will be provided through the
study of dynamically evolving radio spectra, caused by diffractive and
refractive scintillation in the local ISM. Simultaneous observations
from optically thin to optically thick frequencies will provide strong
constraints on any model for the GRB remnant. SKA will also
allow for extremely rapid (within minutes, if not seconds) follow-up
observations by electronic steering of the array.  SKA observations
have the sensitivity to detect GRB afterglows out to redshifts of 10
or greater. They therefore allow studies of the high redshift
universe, and measure the massive star formation (and massive star
death rate!) history of the universe, unbiased by optical
obscuration. Wide field surveys may detect thousands of GRBs within a
few days of observing (depending on the amount of collimation, the
width of the GRB luminosity function, the number of GRBs like GRB
980425/SN\,1998bw). There would be new ones, but many more would be old,
fading afterglows, allowing the determination of the multi-variate
GRB-optical-radio luminosity function of GRBs.  This should elucidate
the relation of the `faint' ones, like GRB\,980425, to the
luminous ones like GRB\,970508.  Such surveys will also tell us
whether GRB fireballs (and which ones) are collimated into jets.  }

\section{Introduction}
Gamma-ray bursts (GRBs) are the strongest phenomenon seen at
$\gamma$-ray wavelengths. Since their discovery in the 1970s these
events, which emit the bulk of their energy in the $0.1 - 1.0$ MeV
range, and whose durations span milliseconds to tens of minutes, posed
one of the great unsolved problems in astrophysics. Until recently, no
counterparts (quiescent as well as transient) could be found and
observations did not provide a direct measurement of their distance,
and thereby the true energy output was unknown by several orders of
magnitude. The breakthrough came in early 1997, when the Wide Field
Cameras aboard the Italian-Dutch BeppoSAX satellite allowed rapid and
accurate localization of GRBs. Follow-up on these positions resulted
in the discovery of X-ray \cite{cfh+97}, optical \cite{vgg+97} and
radio afterglows \cite{fkn+97}. These observations revealed that GRBs
come from `cosmological' distances. GRBs are by far the most luminous
photon sources in the universe, with (isotropic) peak luminosities in
$\gamma$ rays up to $10^{52}$ erg/s, and total energy budgets up to
several $10^{53-54}$ erg (e.g., \cite{kdo+99}). The optical signal from
GRB is regularly seen to be 10 magnitudes brighter (absolute) than the
brightest supernovae, and once even 18 magnitudes
brighter\cite{abb+99}.

Here we discuss the current status of GRB afterglow observations
(Sect. \ref{Fire} to \ref{sec:bea}) and discuss the unique potential
of SKA observations of GRBs (Sect. \ref{sec:ska}).

\section{Relativistic blast-wave models}
\label{Fire}

GRB afterglows are in good agreement with, so called,
fireball-plus-relativistic blast-wave models (see \cite{pir99} for an
extensive review).  The basic model is a point explosion with an
energy of order $10^{52}$ ergs, which leads to a `fireball', an
optically thick radiation-electron-positron plasma with initial energy
much larger than its rest mass that expands
ultra-relativistically. The GRB may be due to a series of `internal
shocks' that develop in the relativistic ejecta before they collide
with the ambient medium. When the fireball runs into the surrounding
medium a `forward shock' ploughs into the medium and heats it, and a
`reverse shock' does the same to the ejecta. As the forward shock is
decelerated by increasing amounts of swept-up material it produces a
slowly fading `afterglow' of X rays, then ultraviolet, optical,
infrared, millimetre, and radio radiation.

\noindent{\bf Confirmation of the relativistic blast-wave model.}
Radio light curves of the afterglow of GRB\,970508 show variability on
time scales of less than a day, but these dampen out after one month
\cite{fkn+97} (see Fig. \ref{fig:radiofrail}). Interpreting this as the
effect of source expansion on the diffractive interstellar
scintillation a source size of roughly 10$^{17}$ cm was derived,
corresponding to a mildly relativistic expansion of the shell
\cite{fkn+97}.

The first X-ray and optical (but see \cite{gtv+99}) afterglows show
power-law temporal decays, with power-law exponents in the range 1 to
2. These afterglow light curves agree well with the predictions of the
relativistic blast-wave model (e.g., \cite{wrm97}).

The broad-band afterglow spectra are also power laws (in four distinct
regions); together with the observed decrease of the cooling break and
the peak frequency the observations conform nicely with simple
relativistic blast-wave models in which the emission is synchrotron
radiation by electrons accelerated in a relativistic shock
\cite{gwb+98,wg99}.

The brightness temperature of the GRB\,990123 optical flash
\cite{abb+99} exceeds the Compton limit of $10^{12}$\,K, confirming the
highly relativistic nature of the GRB source \cite{gbw+99}.

\section{Progenitors and the cause of the explosion}
\label{sec:sne}

The GRB and the afterglow are produced when relativistic ejecta are
slowed down; no observable radiation emerges directly from the `hidden
engine' that powers the GRB. Thus, in spite of all recent discoveries
the origin of GRBs remains unknown (although an important link may be
provided by the possible connection of GRBs to SNe).  Currently
popular models for the origin of GRBs are the neutron star-neutron
star and neutron star-black hole mergers, white dwarf collapse, and
core collapses of very massive stars (`failed' supernovae or
hypernovae). These models can in principle provide the required
energies.

\begin{figure}
\centerline{\psfig{figure=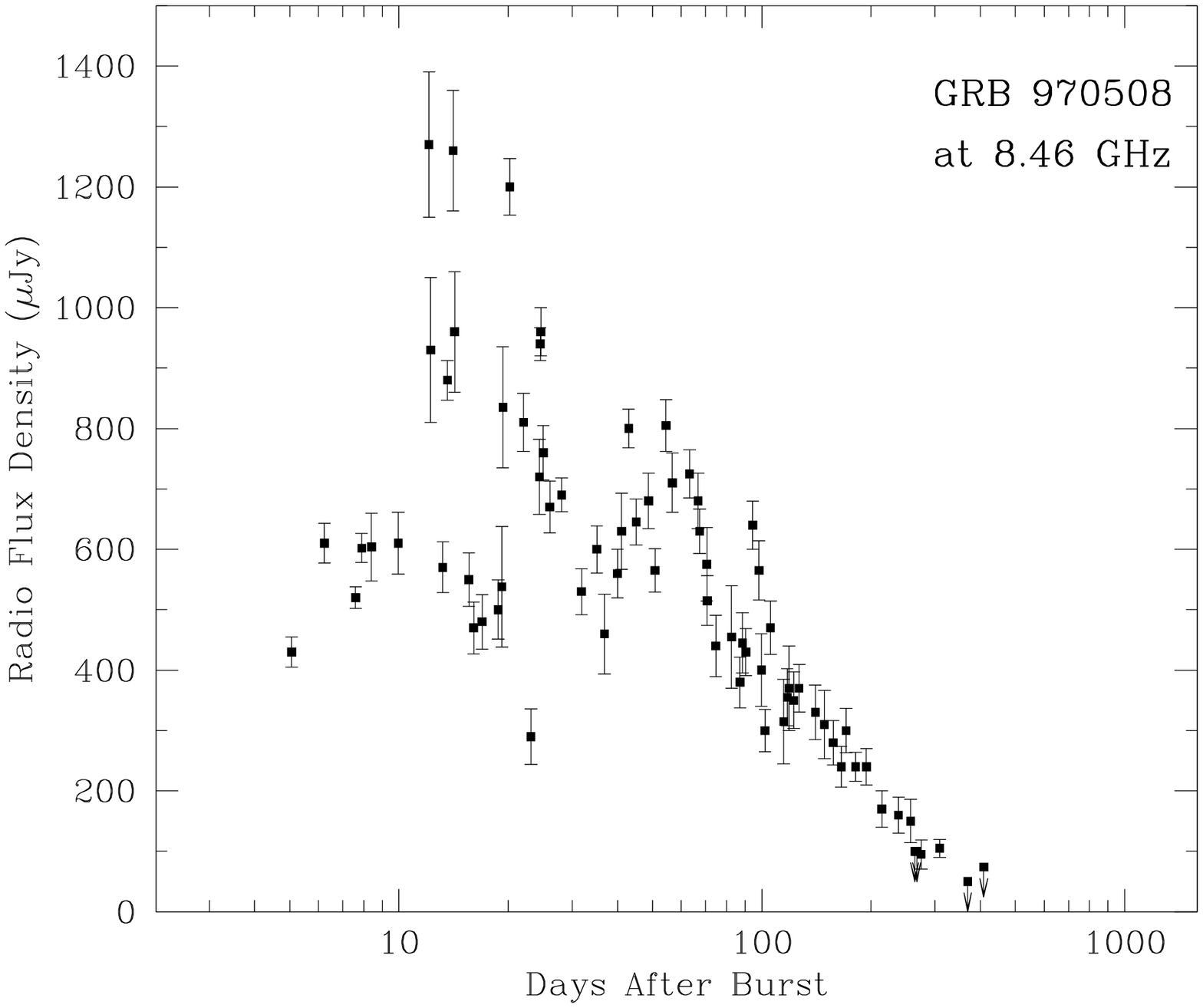,width=8cm}\psfig{figure=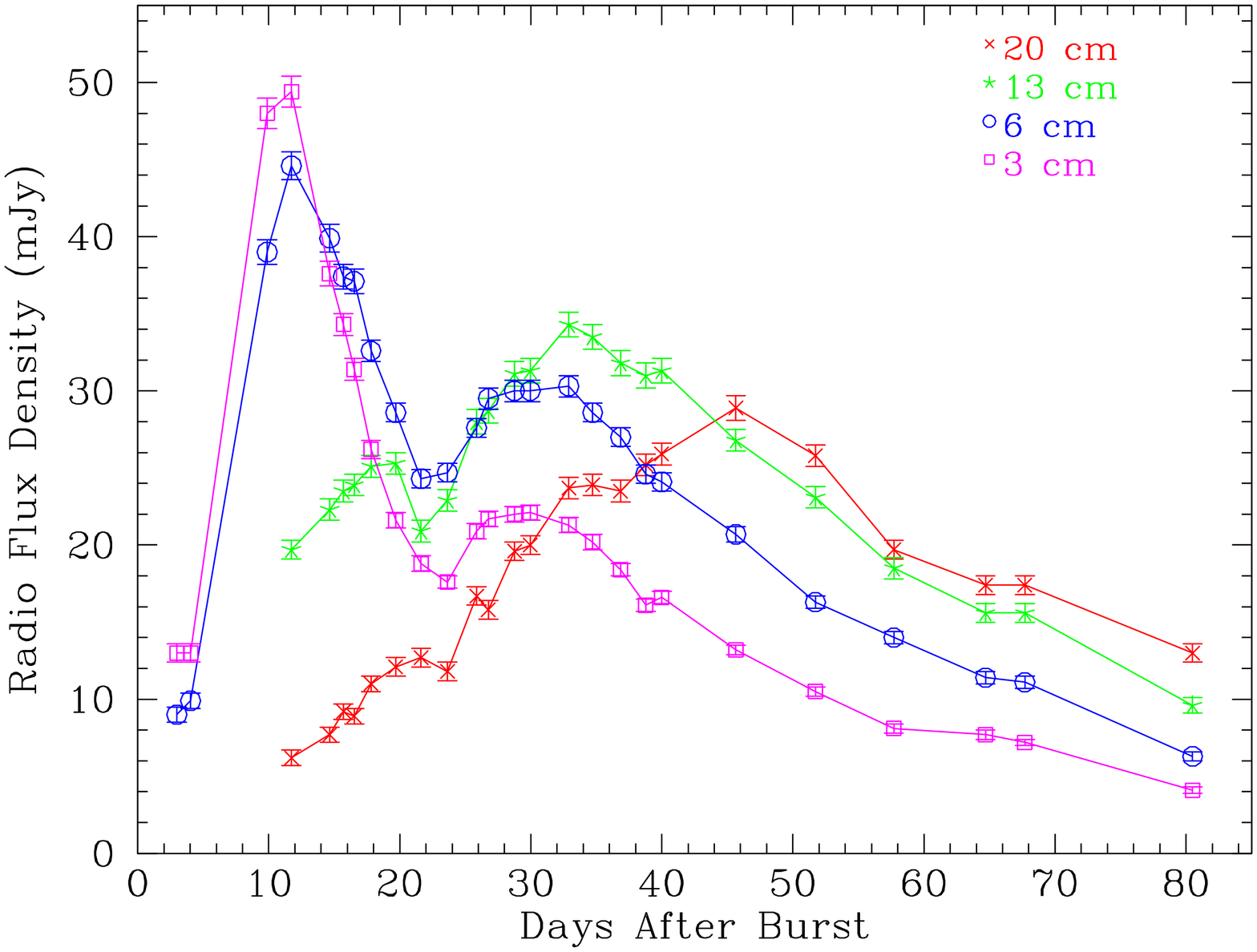,width=8cm,angle=-90}}
    \caption{\small LEFT: Very Large Array observations of GRB\,970508
    (from Frail et al. \cite{fwk99}). The large intensity fluctuations
    during the first 3 weeks are attributed to ISS. The fluctuations
    dampen because of the expansion of the source. RIGHT: The radio
    light curve of SN\,1998bw obtained from the Australia Telescope
    Compact Array (from Kulkarni et al. \cite{kfw+98}).  The
    brightness temperature $T_B>10^{12}$ K in the first month and
    exceeding $10^{13}$ K in the first week.  Since this is greater
    than the $5\times 10^{11}$ K limit for catastrophic inverse
    Compton scattering losses the radio emitting sphere must have
    expanded at relativistic speeds, $\Gamma \gsim 2$.
    \label{fig:radiofrail}}
\end{figure}

\noindent{\bf SN\,1998bw/GRB\,980425.}  Galama et al. \cite{gvv+98}
discovered a relatively rare and bright SN of type Ic within the small
BeppoSAX localization of GRB\,980425, and suggested that the two
objects are connected.  A conservative estimate of the probability of
a chance coincidence of the supernova and the GRB is $9 \times
10^{-5}$ \cite{gvv+98}.  In the radio, the SN rapidly brightened and
became one of the most luminous radio SNe \cite{kfw+98}. Kulkarni et
al. \cite{kfw+98} drew attention to the fact that the radio emitting
shell in SN\,1998bw must be expanding at relativistic velocities,
$\Gamma \gsim 2$; see Fig. \ref{fig:radiofrail}. This relativistic
shock could well have produced the GRB at early times. The consequence
of accepting such an association is that the $\gamma$-ray peak
luminosity of GRB\,980425 and its total $\gamma$-ray energy budget are
much smaller (a factor of $\sim$ 10$^5$) than those of `normal'
GRBs. GRB\,980425 is thus a member of a new class of GRBs -- low
luminosity GRBs related to nearby SNe (SN\,1998bw is at $z = 0.0085$).
Such GRBs may well be the most most frequently occuring GRBs!

\noindent{\bf GRBs and SNe.} But perhaps most, if not all, GRBs are
associated with supernovae. There is growing evidence linking the
usual GRBs -- the cosmologically located GRBs -- to SNe. The most
direct evidence comes from the suggestion of an underlying SN in
the afterglow of GRB\,980326 \cite{bkd+99}.  The SN is revealed readily
by its distinctive UV-poor spectrum against the broad-band afterglow.
Also for GRB\,970228 there is evidence that a supernova dominated the
light curves at late times \cite{rei99,gtv+99}.  So there may not be a
dichotomy between `normal' and supernova GRBs, only a gradual
transition.  The relation between cosmologically located GRBs like
GRBs (980326, 970228) and GRB\,980425/SN\,1998bw is as yet unclear.

\noindent{\bf Collapsar model.}  All these observational developments
support the collapsar model pioneered by Woosley and collaborators
(see \cite{wmh99} and refs therein) in which massive stars core
collapse to form black holes. Energy is somehow extracted from the
spinning black hole and the jets drill their way out and power the
GRBs and their afterglows.

\section{GRBs as potential probes of the high-redshift universe} 
\label{sec:sfr}
Host galaxies have been seen in most optical afterglow images. The
detection of [O~II] $\lambda$ 3727 and Lyman~$\alpha$ emission from
some hosts indicates that these are sites of vigorous star formation.
The observed connection between some GRBs and star forming regions
suggests that GRBs occur at critical phases in the evolution of
massive stars. If GRBs are related to the deaths of massive stars
(whose total lifetime is very short), their rate is proportional to
the star formation rate (SFR).  In that case GRBs may very well be at
very high redshifts, with $z \sim 6$ or greater, for the faintest
bursts (e.g. \cite{wbbn98}). GRBs may therefore become a powerful tool
to probe the far reaches of the universe by guiding us to regions of
very early star formation, and the (proto) galaxies and (proto)
clusters of which they are part. The redshifts determined so far range
between $z=0.41$ and $z = 3.42$.

\section{The early afterglow}
\label{sec:early}
The discovery of a very bright and brief optical flash coincident in
time with GRB\,990123 \cite{abb+99} shows that the early optical
signal from GRB can be some 18 magnitudes brighter than the brightest
supernovae. The reverse shock could cause emission 
that peaks in the optical waveband and is observed only during or just
after the GRB. GRB\,990123 would then be the first burst in which all
three emitting regions have been seen: internal shocks causing the
GRB, the reverse shock causing the prompt optical flash, and the
forward shock causing the afterglow \cite{mr99,sp99b,gbw+99}. 


\section{Strongly anisotropic outflow (beaming)}
\label{sec:bea}
An important uncertainty concerns the possible beaming of the
$\gamma$-ray and afterglow emissions. This has an immediate impact on
the burst energetics, and the nature and number of events needed to
account for the observed burst rate \cite{rho99}. If the afterglow is
beamed with opening angle $\theta$, a change of the light curve slope
occurs at the time when the Lorentz factor $\Gamma$ of the blast wave
equals $1/\theta$. Slightly later the jet begins a lateral expansion,
which causes a further steepening of the light curve.  Perhaps such a
transition has been observed in the optical afterglow light curve of
GRB\,990123 (e.g., \cite{kdo+99}). A similar transition was better
sampled in afterglow data of GRB\,990510; optical observations of
GRB\,990510, show a clear steepening of the rate of decay of the light
between $\sim$ 3 hours and several days \cite{hbf+99}. Together with
radio observations \cite{hbf+99}, which reveal a similar steepening of
the decline, it is found that the transition is very much
frequency-independent; this virtually excludes explanations in terms
of the passage of the cooling or the peak frequency, but is what is
expected in case of beaming. Harrison et al. (1999) derive a jet
opening angle of $\theta = 0.08$ radians, which for this burst would reduce
the total energy in $\gamma$ rays to $\sim 10^{51}$ erg.

\section{The unique potential of SKA observations of GRB afterglows.}
\label{sec:ska}

\begin{itemize}
\item{\bf Interstellar Scintillation.}
As discussed in Sect. \ref{Fire}, the observations of GRB
afterglows are in good agreement with the relativistic-blast wave
model. However, the expansion rate and size of the blast wave have
never been observed directly. Observations of the size, expansion rate
and the shape of the GRB remnant would provide a stringent test of the
relativistic blast wave model. The size $d$ of the GRB remnant is of
the order of

\begin{equation}
  d = \gamma c t,
\label{eq:size}
\end{equation}

where $\gamma$ is the Lorentz factor of the blast wave, $c$ is the
speed of light and $t$ the time in the observer's frame.  Hence, after
1 week the source size is $2 \gamma$ light-weeks, which, at a typical
redshift of $z \sim1$ corresponds to an angular size of $\sim$ 1
microarcsecond. 

Let us take GRB\,970508 as an example.  Radio light curves of the
afterglow of this GRB show rapid variability on time scales of less
than a day, but these dampen out after one month \cite{fkn+97}. The
reduced flux density modulations, interpreted as diffractive
interstellar scintillation (DISS), are caused by the expansion of the
source and then imply an angular diameter of at most a few
$\mu$arcsec.  At the redshift of GRB\,970508 this corresponds to a
linear diameter of roughly 10$^{17}$ cm corresponding to a mildly
relativistic expansion of the shell \cite{fkn+97}.  Similar estimates
of the source size were derived from the observed flux density for
frequencies below the self-absorption frequency \cite{fkn+97}, and
from the presence of several breaks in the spectral energy
distribution of GRB\,970508 \cite{wg99}.  It is clear that with the
current resolution and sensitivity of earth-bound Very Large Baseline
Interferometry it is going to be impossible to ever obtain a direct
measurement of the source size, except for the nearest GRBs like
GRB\,980425/SN\,1998bw.

However, the example of GRB\,970508 shows that indirect source size
estimates of GRB remnants can be obtained by observations of
interstellar scintillation (DISS and RISS).
The current observations are still severely sensitivity limited but
as we discuss below there can be fantastic progress with the 
sensitivity provided by SKA.

Strong scattering can be observed
at frequencies below the transition frequency $\nu_0$ (typically
$\nu_0 \sim$ 5 GHz at high galactic latitudes \cite{wal98}).  
For $\nu>\nu_0$, the scattering is weak and the
modulations scale as $(\nu/\nu_0)^{-17/12}<1$.  For $\nu<\nu_0$ we
will see strong diffractive scintillation only if the size of the
source, $\theta_S<\theta_D=\theta_{F_0}(\nu/\nu_0)^{6/5}$. 
Depending on the properties of the turbulent plasma screen 
we may expect to encounter such conditions in the first day(s) after the 
GRB event. 

The other ISS parameters of interest are the decorrelation timescale $t_{\rm
diff}$ (time for significant changes in the detected flux), and the
bandwidth over which the diffractive ISS is decorrelated, $\Delta\nu =
\nu_0 (\nu/\nu_0)^{22/5}$. These scale as: $t_{\rm diff} \propto
\nu^{1.2}$ and $\Delta \nu_{\rm d} \propto \nu^{4.4}$. As emphasized 
by Goodman \cite{goo97} all these observables carry independent information
on the properties of the ISM.

Observations of the modulation index, decorrelation time scale and
decorrelation bandwidth as a function of frequency, and the
determination of the transition frequency $\nu_0$ between weak and
strong scattering, in early GRB afterglows, hence provide a wealth of
information on the dynamically changing size and shape of the source.
In principle the scintillation tool allows us to infer a crude measure
of the morphology of the radio source: is it ring-like, as in
spherical blast-wave models, or jet-like, in currently popular models.
In the decaying, sub- or non-relativistic phase, the source may
develop double structure (approaching and receding jet components)
leading to distinct patterns in the dynamically changing spectra.

The scintillation method, however, will need a proper calibration
before it will release its potential.  This is clearly shown by the
inferred sizes of two recently discovered scintillating quasars. For
the radio quasar PKS 0405-385 \cite{kjw+97} an angular size of $\sim$
5 $\mu$arcsec was estimated. However, for J1819+3845, Dennett-Thorpe
and de Bruyn \cite{dtb00} estimate a size of $\sim$ 25 $\mu$arcsec at
5 GHz (possibly related to relatively nearby plasma turbulence).  The
calibration of scintillation screen properties can be provided by the
observation of angularly nearby pulsars, the perfect point sources, of
which SKA could easily detect about 1 per square degree.


SKA will provide the right technical specifications for these exciting
and unique observations. The proposed instantaneous bandwidth: $0.5 +
f/5$ GHz and the large number of spectral channels: 10$^4$, allow the
recording of dynamic spectra of GRBs, as is now common for pulsars.

\item{\bf Synchrotron self absorption.} The large instantaneous
bandwidth, the large number of spectral channels, and sensitiviy of
SKA will also allow detailed study of the transition from optically
thin to optically thick frequencies and the evolution of the shape and
location of this transition. Such high quality observations will
provide strong constraints on models of GRB afterglows.

\item{\bf Supernova-GRBs.}  As discussed in Sect. \ref{sec:sne} the
most common GRBs may very well be the low luminosity GRBs like GRB
980425/SN\,1998bw. The bright radio emission of such sources may
easily be detected out to redshifts of $z \sim 1$, with SKA's
sensitivity.  From the duration of the radio phase in SN\,1998bw, its
typical brightness of $\sim$ 10 mJy at GHz frequencies, and some
simple order-of-magnitude extrapolations to the whole sky, we derive
that at any given time there will be several tens of such fading
supernova-GRB radio afterglows above a flux density of 1 $\mu$Jy per square
degree!  Most of these sources will be distant, hence small and
scintillating.  This is how they could be discerned. What
distinguishes them from AGN, however, is that they would appear at
places where previously there would have been no radio source.  A
survey of the sky, carried out with SKA in its first year, could
provide the template, against which to pick up these new sources.
(Nota bene, this is also how new radio SNe would be discovered, but
they are typically orders of magnitude fainter and rarely reach the
radio brightness temperatures of GRB afterglows).

\item{\bf Rapid response.} The early radio afterglow emission of GRBs
is currently hard to observe: the sources are very faint at ages less
than 1 day and the current response time, which is typically a few
hours, is too slow. Soon, with the launch of the gamma-ray satellite
HETE-II, the response time may be very much improved upon (positions
will be available within tens of seconds of the event), but the
sensitivity is expected to remain problematic for such early
observations. SKA will provide the required sensitivity plus it will
have this other unique capability: the possibility of very rapid
response (we may, however, have to build a few SKA's to cover both the
northern and southern hemispheres, and provide 24h watch!).  We
envision that SKA may be triggered directly by future gamma-ray
spacecraft, and then rapidly electronically steer to the location on
the sky. Such observations may provide insight in the physics of the
fireball at very early times, for example we may expect to detect
emission from the reverse shock \cite{sp99a}.

\item{\bf Wide Field Surveys.} SKA will be a unique instrument for
surveying large areas of sky (as noted above in connection with the
GRB\,980425/SN\,1988bw association). If the GRB luminosity function is
bimodal (i.e. GRBs like 980425 versus the more distant GRBs like
970508) then such surveys may be dominated by supernova-GRBs. However,
a substantial number of the more distant GRBs may be discovered too,
depending on the amount of collimation into jets. If the GRB
luminosity function is not bimodal, but very wide then we may also
detect a substantial number of radio afterglows from intermediate
luminosity GRBs ($E_\gamma \sim 10^{48-50}$ erg).  Thus, thousands of GRBs
and supernova-GRBs may be discovered from such surveys by their
distinct observational characteristics (for example a self-absorbed
synchrotron spectrum, observed ISS and the characteristic damping of
ISS fluctuations with time due to expansion of the source, the fact
that GRB afterglows are not expected to be recurrent, etc). An
important aspect of a radio selected survey is its unbiased-to-dust
nature.  Afterglows may thus be discovered independent of an optical
or even a GRB identification.

The statistics of radio afterglows may be compared with the numbers
expected from specific spherical- or jet-fireball models. As the bulk
Lorentz factor, $\Gamma$ decreases with time after the event, the
observer sees more and more of the emitting surface, $\theta \sim
1/\Gamma$. It follows that if gamma ray bursts are highly collimated,
many more radio transients should be observed without associated gamma
rays than with them. The ratio of expected (assuming spherical
symmetry) to observed number of afterglows is thus a direct measure of
the amount of collimation in GRB fireballs.

The number of radio afterglows may also be compared to that of optical
afterglows (subject to obscuration by dust) to reveal possibly dusty
environments (expected for massive star progenitor models).

\item{\bf The high-$z$ universe and the cosmic star formation/death 
history.}  A GRB at a redshift $z \sim 1$ may
easily be a mJy bright. With SKA's senstivity such radio counterparts
can be detected out to redshifts of 10 or greater.  SKA will also be
sufficiently sensitive that the radio emission of GRB hosts can be
studied. Currently this is barely feasible. Such observations may
provide information on the progenitors of GRBs (e.g., young or old
stellar populations). Also, as discussed in Sect. \ref{sec:sfr} GRBs
are expected to trace the star formation rate (SFR) in the
universe. Vice versa by observing GRB hosts we will learn about the
star formation (and star death!) history.  
Observations in the (sub-)mm band suggest
that star formation at high redshift is dominated by dusty
star burst galaxies \cite{bsik99}. Estimates of the SFR of GRB host
galaxies can be accurately determined by radio observations.
These observations will be insensitive to dust, a crucial fact if one
wants to be complete.
\end{itemize}

\newpage

\end{document}